\documentclass[aps,pra,twocolumn,superscriptaddress, showkeys]{revtex4-1}

\usepackage{mathptmx}
\usepackage{graphicx}
\usepackage{physics}            
\usepackage{amsfonts}
\usepackage{amssymb}            
\usepackage{braket}
\usepackage{amsmath}
\usepackage{float}
\usepackage{verbatim}
\usepackage{nomencl}
\usepackage{hyperref}
\usepackage{xcolor}
\hypersetup{
	colorlinks   = true, 
	urlcolor     = blue, 
	linkcolor    = blue, 
	citecolor   = blue 
}

\begin{document}


\title{Entanglement of indistinguishable particles: a comparative study}


\author{Ant\^onio C. Louren\c{c}o }

\email[]{lourenco.antonio.c@gmail.com}
\affiliation{Departamento de F\'isica, Universidade Federal de Santa Catarina, CEP 88040-900, Florian\'opolis, SC, Brazil }

\author{ Tiago Debarba
}
\affiliation{Universidade Tecnol\'ogica Federal do Paran\'a, Campus Corn\'elio Proc\'opio, Avenida Alberto Carazzai 1640,
Corn\'elio Proc\'opio, Paran\'a 86300-000, Brazil}

\author{Eduardo I. Duzzioni
}
\affiliation{Departamento de F\'isica, Universidade Federal de Santa Catarina, CEP 88040-900, Florian\'opolis, SC, Brazil }


\date{\today}

\begin{abstract}
There is a debate in course about the quantification of entanglement of indistinguishable particles and recently a new method due to Lo Franco and Compagno (LFC) [Sci. Rep. \textbf{6}, 20603 (2016)] appeared with the aim of settling the controversies of this debate. In this work we show that the LFC approach can be recovered using the second quantization formalism and a local single-particle basis to perform the trace over one particle state. The use of the second quantization formalism has the advantage of being well known in many-body physics. We also compare the LFC and second quantization approaches, or equivalently, the effect of the partial trace over one particle depending on the local and nonlocal caracter of the single-particle basis. Additionally, we study the role played by the exchange correlations in the quantification of entanglement of indistinguishable particles. 
\end{abstract}


\maketitle

\section{Introduction}
\label{intro}
Entanglement is a unique characteristic  of quantum mechanics and, maybe, it is considered what makes the world of quantum mechanics so different from the classical one. The idea of entanglement was born with E. Schr\"odinger \cite{schrodinger1935gegenwartige} and used by A. Einstein, B. Podolsky, and N. Rosen as a tentative to show that quantum mechanics was incomplete \cite{EPR1935}. The first direct experimental measure of bipartite entanglement was reported in \cite{walborn2006experimental}. The interest in the study of entanglement has increased since one noted that it could be used in some quantum processes, as quantum computation \cite{jozsa2003role}, quantum cryptography \cite{aktas2016}, quantum teleportation \cite{wang2015,pirandola2015}, and superdense coding \cite{feng2016}. 

Entanglement is also used in some processes of quantum computing involving systems of indistinguishable particles like quantum dots \cite{tan2015} and Bose-Einstein condensates in optical lattices \cite{anderlini2007}. Therefore, it is necessary to have a well-established theory to deal with entanglement of indistinguishable particles. According to the mathematical representation of a state of indistinguishable particles commonly used in texts books and many-body theories, in which it is necessary to symmetrize the multi-particle state, the system \textit{appear} to be entangled, generating discussions if a system with only exchange correlations is entangled or not. Because of that, there is a debate in course about the determination of entanglement in systems with exchange correlations \cite{schliemann2001,li2001,pavskauskas2001,eckert2002,ghirardi2002,gittings2002,wiseman2003,ghirardi2004,plastino2009,iemini2013,franco2016,benatti2017,franco2018indistinguishability}. 

In order to exclude exchange correlations from states of indistinguishable particles, a method named Slater rank has been developed for the characterization of entanglement \cite{schliemann2001,li2001}. In this method, the Slater-Schmidt decomposition, analogously to Schmidt decomposition, determines that  states possessing a single Slater determinant or a convex combination of them are separable \cite{schliemann2001,eckert2002,pavskauskas2001}. After applying the Slater decomposition one can quantify entanglement using the von Neumann entropy \cite{li2001,pavskauskas2001}. The results presented in Ref. \cite{pavskauskas2001} are similar to those presented in \cite{plastino2009,iemini2013}, but the equality between them is achieved through a shift in the von Neumann entropy by the quantity $\ln N$ in the former case, where $N$ is the number of particles in the system. To capture correlations beyond entanglement and exchange correlations of general mixed states, see for instance, the quantumness of correlations of indistinguishable particles \cite{iemini14,debarba17}.

Recently, a method due to Lo Franco and Compagno (LFC) appeared with the purpose of solving the debate on entanglement of indistinguishable particles \cite{franco2016}. In this new approach \cite{sciara2017,franco2018indistinguishability,bellomo2017,compagno2018dealing}, there is no need of labeling indistinguishable particles due to symmetrization by permutation of particles. In a later study \cite{sciara2017}, in which the authors extend their method to Schmidt decomposition of indistinguishable particles, it is stablished a comparison between their method and some aspects of the second quantization approach \cite{li2001,pavskauskas2001}. In ref. \cite{benatti2017} is made a comparison between the LFC approach and an approach based on mean values of commuting observables to verify that the notion of entanglement is not absolute, but depends on the states and the choice of the observables. In this work we show how to recover the results of LFC \cite{franco2016} through the application of the second quantization formalism and the notion of localized partial trace. As application we explore the entanglement of two indistinguishable particles trapped in an asymmetric double-well system. The effect of the nonlocal single-particle basis to take the trace over one particle and the role played by the exchange correlations \cite{plastino2009,iemini2013} are analyzed and compared to the LFC approach.

This paper is organized as follows: in the Sec. \ref{sec:LFC} we review the LFC approach and in the Sec. \ref{sec:RLR} we show how to recover the results of LFC's method from the second quantization approach. The differences between the LFC and the second quantization approaches for a system composed of two identical particles trapped in an asymmetric double-well are analized in Sec. \ref{sec:App}. In Sec. \ref{sec:Dis} we discuss the difference between these two measures of entanglement of indistinguishable particles.
\section{Lo Franco and Compagno's Method}
\label{sec:LFC}

Lo Franco and Compagno developed an approach to deal with indistinguishable particles in which it is not necessary to use the symmetrization or antisymmetrization of the states of many particles \cite{franco2016}. For example, two indistinguishable particles in states labeled by $\phi$ and $\psi$ are treated as a single entity $\ket{\phi,\psi}$.
This way of representing the state of two indistinguishable particles is interesting provided that its notation does not induce naturally the reader to consider it entangled, as occurs with the usual notation of two-particle state due to the symmetrization postulate.  
Instead, what is symmetrized is the inner product, which for two particles has the form,
\begin{equation}
\label{eq:pis}
\braket{\varphi,\zeta|\phi,\psi} \equiv \braket{\varphi|\phi} \braket{\zeta|\psi} + \eta\braket{\varphi|\psi}\braket{\zeta|\phi},
\end{equation}
\noindent with $\eta = 1$ for bosons and $\eta = -1 $ for fermions. 
Thus, if the probability amplitude of finding two particles in the same state is measured, one has $\braket {\varphi, \varphi|\phi,\psi} = (1+\eta) \braket {\varphi | \phi} \braket {\varphi | \psi} $. For fermions, the probability amplitude is zero, obeying the Pauli exclusion principle, and for bosons, it has maximum value.

From symmetry and linearity of one-particle states in equation \ref{eq:pis}, one can write the symmetric inner product of states with spaces of different dimensionality. Then, considering 
$\ket{\tilde{\mathrm{\Phi}}}=\ket{\varphi_{1}, \varphi_{2}}$ the unnormalized state of two identical particles, the inner product with a single-particle state $\ket{\psi_k}$ is
\begin{equation} 
\label{eq:inner_prod_1}
\begin{split}
\braket{\psi_{k}|\cdot|\varphi_{1}, \varphi_{2}}\equiv \braket{\psi_{k}|\varphi_{1}, \varphi_{2}}=& \braket{\psi_{k}|\varphi_{1}}\ket{ \varphi_{2}}\\
&+\eta\braket{\psi_{k}|\varphi_{2}}\ket{ \varphi_{1}}.
\end{split}
\end{equation}
\noindent This is a projective measurement on a single particle, where the unnormalized two particle state is projected onto $\ket{\psi_{k}}$. Therefore, the normalized reduced pure-state of a single particle is 
\begin{equation}
\ket{\phi_{k}}=\frac{\braket{\psi_{k}|\mathrm{\Phi}}}{\sqrt{\braket{\mathrm{\Pi}_{k}^{(1)}}_{\mathrm{\Phi}}}},
\end{equation}
\noindent where $\braket{\psi_{k}|\mathrm{\Phi}}$ is obtained from Eq. (\ref{eq:inner_prod_1}) and $\mathrm{\Pi}_{k}^{(1)}=\ket{\psi_{k}}\bra{\psi_{k}}$ is an one-particle projection operator. Notice that $\ket{\mathrm{\Phi}}=\frac{1}{\sqrt{\mathcal{N}}}\ket{\tilde{\mathrm{\Phi}}}$ is the normalized two-particle state, with $\mathcal{N}=1+\eta|\braket{\varphi_{1}|\varphi_{2}}|^2$.

Defining the one-particle identity operator as $\mathbb{I}^{(1)}=\sum_{k}\mathrm{\Pi}_{k}^{(1)}$ and using the linearity property of the projection operators,  
\begin{equation} 
\label{eq:ext_prod_2}
\begin{split}
\ket{\psi_{k}}\bra{\psi_{k}} \cdot \ket{\varphi_{1}, \varphi_{2}} \equiv& \braket{\psi_{k}|\varphi_{1}} \ket{\psi_k, \varphi_{2}} \\
&+\eta \braket{\psi_{k}|\varphi_{2}} \ket{\varphi_{1}, \psi_k},
\end{split}
\end{equation}
we obtain, 
\begin{equation}\label{eq:identity}
\mathbb{I}^{(1)}\ket{\mathrm{\Phi}}=2\ket{\mathrm{\Phi}},
\end{equation}
so that the probability of obtaining the state $\ket{\psi_k}$ is $p_{k}=\braket{\mathrm{\Pi}_{k}^{(1)}}_{\Phi}/2$.

The partial trace in the LFC approach differs from the usual way of obtaining the partial trace of distinguishable particles only by the inner product. In such approach the partial trace is taken as follows 
\begin{equation}
\Tr_{(1)}\ket{\upPhi}\bra{\upPhi}=\sum_{k}\braket{\psi_{k}|\upPhi}\braket{\upPhi|\psi_{k}},
\end{equation}
with the normalized reduced density matrix given by		
\begin{equation}
\rho_{1}=\frac{1}{2}\Tr_{(1)}\ket{\upPhi}\bra{\upPhi}=\sum_{k}p_k \ket{\phi_{k}}\bra{\phi_{k}}.
\end{equation}
The factor $1/2$ appears in the above equation due to condition (\ref{eq:identity}).				

In the LFC approach the authors define a local measurement on one particle as the one performed on a region of space $M$ (site or spacial mode), where the particle has nonnull probability of being found \cite{franco2016}. 
According to this definition, the \textit{localized partial trace}, i.e., that one taken over a single-particle local basis, is somewhat different from its usual form, which one is taken over a complete basis involving all degrees of freedom of a single particle. When we are dealing with indistinguishable particles, it is called nonlocal single-particle basis \cite{sciara2017}. Therefore, performing the localized partial trace, we obtain the reduced density matrix
\begin{equation}
\rho^{(1)}_{M}=(1/\mathcal{N}_M)\Tr^{(1)}_{M}\ket{\upPhi}\bra{\upPhi},
\end{equation}
\noindent where $\mathcal{N}_M$ is a normalization constant such that $\Tr^{(1)}\rho_{M}^{(1)}=1.$ 
In order to quantify entanglement between two particles, it is used the entanglement entropy 
\begin{equation}
E_{M}(\ket{\upPhi}) \equiv S(\rho_{M}^{(1)})=-\sum_{i}\lambda_{i}\ln\lambda_{i},
\end{equation}
\noindent where $S(\rho)=-\Tr(\rho \ln\rho)$ is the von Neumann entropy and $\lambda_{i}$ are the eigenvalues of $\rho^{M}_{(1)}$.
\section{Recovering the LFC results}
\label{sec:RLR}

First we apply the LFC method to a general pure state of two identical particles to obtain the reduced density matrix of one particle. Then we compare it with the reduced density matrix obtained through the second quantization formalism and partial trace on a local single-particle basis. Once the reduced density matrices are equal, both have the same spectrum and consequently the same amount of entanglement.

\subsection{LFC's method}

Let us consider, in accordance with the LFC's method, a general state describing two identical particles
\begin{equation}
\label{eq:GE}
\ket{\Psi}=\sum_{\alpha,\beta, i, j}^{}C_{\beta j}^{\alpha i}\ket{\alpha i, \beta j},
\end{equation}
where Greek symbols ($\alpha, \beta$) refer to spatial labels while Roman letters ($i,j$) refer to complementary degrees of freedom of a particle, as spin, for instance. The density matrix for this state is
\begin{equation}
\rho=\ket{\Psi}\bra{\Psi}=\sum_{\alpha,\beta, i, j}^{}\sum_{\gamma,\phi, k, l}^{}C_{\beta j}^{\alpha i}C_{\gamma k}^{\phi l*}\ket{\alpha i, \beta j}\bra{\gamma k,\phi l}.
\end{equation}
Without loss of generality, we will assume that all labels used to identify the orthogonal modes of a particle are finite, $\alpha, \beta, \gamma, \phi=\{\xi_{1}, \xi_{2}, ..., \xi_{s}\}$ and $i,j,k,l=\{z_{1}, z_{2}, ...,z_{t}\}$. Performing the localized partial trace in only one region of space, say $\xi_1$, we obtain the nonnormalized density matrix
\begin{equation}
\label{eq:RDM1}
\begin{split}
\tilde{\rho}^{(1)}_M\!=\!\Tr_{M}(\rho)\!=\!&\sum_{m=z_1}^{z_t}\bra{\xi_{1}m}\rho\ket{ \xi_{1}m}\!=\!\sum_{m=z_1}^{z_t}\sum_{\alpha, \beta,i,j}^{}\sum_{\gamma, \phi, k, l}^{} \\
& C_{\beta j}^{\alpha i}C_{\gamma k}^{\phi l*} \left( \braket{\xi_{1}m|\alpha i}\braket{\phi l|\xi_{1}m}\ket{\beta j}\bra{\gamma k}\right. \\ 
&\left.+ \eta \braket{\xi_{1}m|\alpha i}\braket{\gamma k|\xi_{1}m}\ket{\beta j}\bra{\phi l}\right. \\
&\left.+ \eta \braket{\xi_{1}m|\beta j}\braket{\phi l|\xi_{1}m}\ket{\alpha i}\bra{\gamma k}\right.\\ 
&\left.+ \braket{\xi_{1}m|\beta j}\braket{\gamma k|\xi_{1}m}\ket{\alpha i}\bra{\phi l}\right)\\
&=\sum_{m=z_1}^{z_t}\left(\sum_{\beta,j}^{}\sum_{\gamma, k}^{}C_{\beta j}^{\xi_{1} m}C_{\gamma k}^{\xi_{1} m*}\ket{\beta j}\bra{\gamma k}\right.\\
&\left. +\eta \sum_{\beta,j}^{}\sum_{\phi, l}^{} C_{\beta j}^{\xi_{1} m}C_{\xi_{1} m}^{\phi l*}\ket{\beta j}\bra{\phi l} \right.\\
&\left. +\eta \sum_{\alpha,i}^{}\sum_{\gamma, k}^{} C_{\xi_{1} m}^{\alpha i}C_{\gamma k}^{\xi_{1} m*}\ket{\alpha i}\bra{\gamma k}\right.\\
&\left. +\sum_{\alpha,i}^{}\sum_{\phi, l}^{}C_{\xi_{1} m}^{\alpha i}C_{\xi_{1} m}^{\phi l*}\ket{\alpha i}\bra{\phi l}\right).
\end{split}
\end{equation}

\noindent In the above equation we have used Eq. (\ref{eq:pis}) to evaluate the inner product. Observe that for our purposes it is enough to compare the reduced density matrices in both methods.

\subsection{Second quantization formalism}
\label{sec:EoP}

Inspired by some methods that use the second quantization formalism to calculate entanglement between indistinguishable particles  \cite{plastino2009,iemini2013}, we show how to recover the LFC results through this well known formalism used in many-bodies physics.  Defining $\mathcal{F}_N^L$ as the Fock space of $N$ particles in which each one occupies $L$ modes, a general state $\ket{\psi}\in \mathcal{F}_N^L$ can be written as
\begin{equation}
\label{eq:PS}
\ket{\psi}=\sum_{{\phi_1,\ldots,\phi_N}=1}^{L}c_{\phi_1,\ldots,\phi_N}a_{\phi_{1}}^{\dagger}...a_{\phi_{N}}^{\dagger}\ket{0},
\end{equation}
\noindent where $a_{\phi_{j}}^{\dagger}$ are bosonic or fermionic creation operators of the orthonormal modes $\phi_{{j}}$ and $\ket{0}$ is the vacuum state.  
To calculate the entanglement between $m$ and $N-m$ particles, one takes the trace of $m$ particles as follows
\begin{equation}
\tilde{\rho}_{N-m}=\Tr_{1}...\Tr_{m}(\ket{\psi}\bra{\psi}),
\end{equation}
\noindent with $\Tr_{1}$ being the trace over one particle and $\Tr_{1}...\Tr_{m}$ over $m$ particles. Notice that $\tilde{\rho}_{N-m}$ is a nonnormalized density matrix and we do not  know which particle is being traced, since they are indistinguishable. The trace we are referring to here is taken over the complete basis of all degrees of freedom of one particle \cite{plastino2009,iemini2013}, as shown below 
\begin{equation}
\label{eq:TEP}
\Tr_{1}(\ket{\psi}\bra{\psi})=\sum_{\phi_j=1}^{L}\bra{0}a_{\phi_{j}}\ket{\psi}\bra{\psi}a_{\phi_{j}}^{\dagger}\ket{0}.
\end{equation}
However, it can be changed depending on our purposes, as will be clear below.\\

To apply this formalism to the general pure state of two particles (\ref{eq:GE}), first we write the density matrix in the second quantization formalism as, 
\begin{equation}
\label{eq:MGE}
\rho=\ket{\Psi}\bra{\Psi}=\sum_{\alpha,\beta, i, j}^{}\sum_{\gamma,\phi, k, l}^{}C_{\beta j}^{\alpha i}C_{\gamma k}^{\phi l*}a_{\alpha i}^{\dagger}a_{\beta j}^{\dagger}\ket{0}\bra{0}a_{\gamma k}a_{\phi l}.
\end{equation}
In order to compare with the LFC approach, we can take the localized partial trace in Eq. (\ref{eq:MGE}), i.e., to apply the procedure (\ref{eq:TEP}) restricted to spatial modes. Thus, the nonnormalized reduced density matrix for one particle is 
\begin{equation}
\label{eq:RDM2}
\begin{split}
\tilde{\rho}_{1}=&\sum_{m=z_1}^{z_t}\bra{0}a_{\xi_{1}m}\rho a_{\xi_{1}m}^{\dagger}\ket{0}\\
&=\sum_{m=z_1}^{z_t}\left[\sum_{\alpha,\beta, i, j}^{}\sum_{\gamma,\phi, k, l}^{}C_{\beta j}^{\alpha i}C_{\gamma k}^{\phi l*}\right.\\
&\left.(\delta_{\xi_{1}\alpha}\delta_{mi}a_{\beta j}^{\dagger}+\eta \delta_{\xi_{1}\beta}\delta_{mj}a_{\alpha i}^{\dagger})\ket{0} \right.\\
&\left.\bra{0}(\delta_{\xi_{1}\phi}\delta_{ml}a_{\gamma k} +\eta \delta_{\xi_{1}\gamma}\delta_{mk}a_{\phi l})\right]\\
&=\sum_{m=z_1}^{z_t}\left(\sum_{\beta, j}^{}\sum_{\gamma, k}^{}C_{\beta j}^{\xi_{1} m}C_{\gamma k}^{\xi_{1} m*}a_{\beta j}^{\dagger}\ket{0}\bra{0}a_{\gamma k}\right.\\
&\left. + \eta \sum_{\beta, j}^{}\sum_{\phi, l}^{}C_{\beta j}^{\xi_{1} m}C_{\xi_{1} m}^{\phi l*}a_{\beta j}^{\dagger}\ket{0}\bra{0}a_{\phi l}\right.\\
&\left. + \eta \sum_{\alpha,i}^{}\sum_{\gamma, k}^{} C_{\xi_{1} m}^{\alpha i}C_{\gamma k}^{\xi_{1} m*}a_{\alpha i}^{\dagger}\ket{0}\bra{0}a_{\gamma k}\right.\\
&\left. +\sum_{\alpha,i}^{}\sum_{\phi, l}^{}C_{\xi_{1} m}^{\alpha i}C_{\xi_{1} m}^{\phi l*}a_{\alpha i}^{\dagger}\ket{0}\bra{0}a_{\phi l}\right), 
\end{split}
\end{equation}
where we used the commutation relation $[a_{\omega q}\!,\!a_{\pi r}^{\dagger}]_{\eta}\!=\!a_{\omega q}a_{\pi r}^{\dagger}$    $-\eta a_{\pi r}^{\dagger}a_{\omega q}=\delta_{\omega\pi}\delta_{qr}$. As can be seen, the reduced density matrices (\ref{eq:RDM1}) and (\ref{eq:RDM2}) are equal. Therefore, it is enough to use the von Neumman entropy of the density matrix (\ref{eq:RDM2}) normalized to obtain the same expression of LFC \cite{franco2016} for the entanglement of particles.   
\section{Application}
\label{sec:App}
\begin{figure}
	\includegraphics[width=0.48\textwidth]{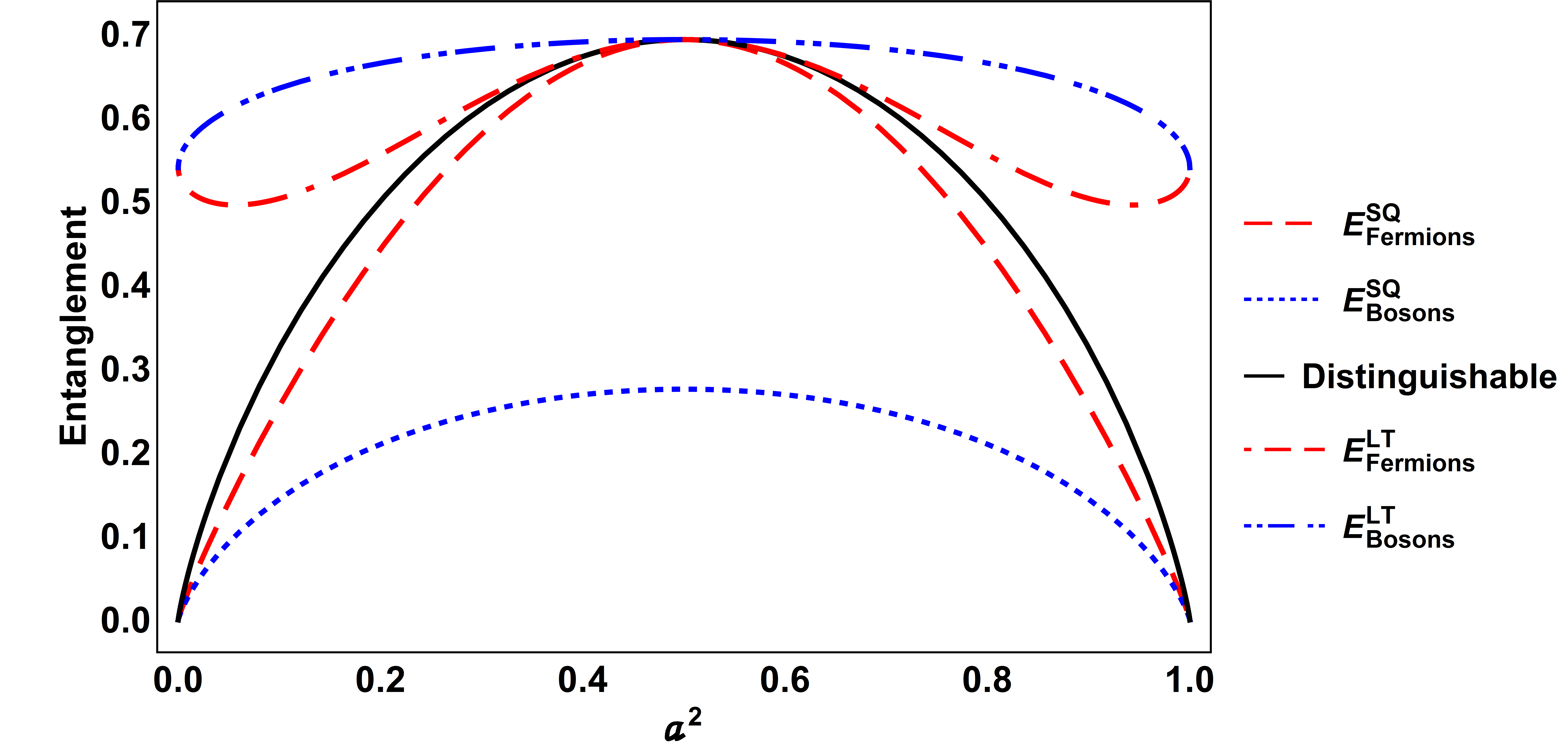}
	
	\caption{Entanglement of bosons, fermions, and distinguishable particles according two different approaches as function of the dimensionless parameter $a^{2}$ with $\theta=0$. In one case the entanglement $E^{LT}(\ket{\Psi})$ between bosons (blue dotted-dotted-dashed line) and fermions (red dotted-dashed-dashed line) is obtained by performing the trace of one particle only in the left well, while in the other case the entanglement $E^{SQ}(\ket{\Psi})$ between bosons (blue dotted line) and fermions (red dashed line) is obtained tracing over a complete basis, i.e., taking the trace over all degrees of freedom of a particle. In both cases the entanglement between bosons and fermions was calculated for the overlap parameter $\chi=0.3$. When $\chi=0$ (black continuous line) the particles are spatially separated and become distinguishable so that both entanglement measures coincide.}
	\label{fig:1}       
\end{figure}

At follows we explore an example already debugged in Ref. \cite{franco2016} to understand the role played by the localized partial trace and particle statistics \cite{cunden2014spatial}. The system is composed by two particles in an asymmetric double well represented by the following state,
\begin{equation}
\label{eq:SDW}
\ket{\psi}= a\ket{L\uparrow, B\downarrow}+ be^{i\theta}\ket{L\downarrow, B\uparrow},
\end{equation}
\noindent with $\ket{B}=\chi \ket{L}+\sqrt{1-\chi^{2}}\ket{R}$, $b=\sqrt{1-a^{2}}$, $\chi=\braket{L|B}$, $0\le \chi \le 1$, and $\braket{L|R}=0$. $L$ and $R$ describe the spatial modes of the left and right wells, respectively. Furthermore, these particles have spin $1/2$ represented by the states $\{\ket{\uparrow},\ket{\downarrow}\}$. It is important to observe that the state (\ref{eq:SDW}) is represented according to Ref.  \cite{franco2016}, i.e., the symmetrization postulate has not been used in such description. This state describes one particle localized in the left well while the other one can tunnel from the right to the left well, but the contrary does not happen. The state (\ref{eq:SDW}) can be rewritten in a more detailed way as 
\begin{equation} 
\label{eq:LFCa}
\begin{split}
\ket{\psi}=&\left(a + \eta b e^{i\theta}\right)\chi\ket{L\uparrow, L\downarrow}+ \\ 
&a\sqrt{1-\chi^2}\ket{L\uparrow, R\downarrow} +b e^{i\theta}\sqrt{1-\chi^2}\ket{L\downarrow, R\uparrow}.
\end{split}
\end{equation}
To quantify the entanglement of particles according to the second quantization formalism, we define first the single particle basis in $\mathcal{F}_{1}^{4}$ as $ \{\ket{L\uparrow},\ket{L\downarrow},\ket{R\uparrow},\ket{R\downarrow}\}$. Then, writing the state (\ref{eq:LFCa}) as specified by this this formalism, we obtain 
\begin{equation} 
\label{eq:EoP}
\begin{split}
\ket{\psi}&=\left(a +\eta be^{i\theta}\right)\chi a_{L\uparrow}^{\dagger}a_{L\downarrow}^{\dagger}\ket{0}+ 
a\sqrt{1-\chi^2}a_{L\uparrow}^{\dagger}a_{R\downarrow}^{\dagger}\ket{0}\\
&+be^{i\theta}\sqrt{1-\chi^2}a_{L\downarrow}^{\dagger}a_{R\uparrow}^{\dagger}\ket{0}.
\end{split}
\end{equation}
To recover the results presented in ref. \cite{franco2016} we perform the localized partial trace on the left well. This procedure is implemented by using Eq. (\ref{eq:TEP}) with the restriction of taking the trace over the left spatial modes $\{ a_{L\uparrow}\ket{0},a_{L\downarrow}\ket{0} \}$. Then, the normalized reduced density matrix is
\begin{equation}
\label{eq:MRLC}
\rho_1=\frac{1}{\mathcal{N}_1}\left(
\begin{array}{cccc}
c_{1} & 0 & c_{4} & 0 \\
0 & c_{1} & 0 & c_{5} \\
c_{4}^{*} & 0 & c_{2} & 0 \\
0 & c_{5}^{*} & 0 & c_{3} \\ 
\end{array}
\right),
\end{equation}
where
\begin{equation}
\begin{split}
&c_{1}=\chi(1+2ab\eta\cos\theta),\\
&c_{2}=b^2 (1-\chi), \\
&c_{3}=a^2 (1-\chi), \\
&c_{4}=\eta b e^{-i \theta} (a+\eta b e^{i\theta}) \sqrt{(1-\chi) \chi}, \\
&c_{5}=a \left(a+\eta b e^{i \theta}\right) \sqrt{(1-\chi) \chi},\\
&\mathcal{N}_1\!=2\! c_{1}\!+\! c_{2}\!+\!c_{3}\!=\!1\!+\!\chi(1\!+\!4\eta ab\cos\theta).
\end{split}
\end{equation}
We highlight that the reduced density matrix (\ref{eq:MRLC}) is the same one as obtained in Ref. \cite{franco2016}. The entanglement between the particles is measured by the von Neumann entropy of the density matrix (\ref{eq:MRLC}) 
\begin{equation}
E^{LT}(\ket{\Psi})=-\Tr \left( \rho_1\ln \rho_1 \right)=-\sum_{i=1}^4 \lambda_i\ln \lambda_i, 
\end{equation}
whose eigenvalues are expressed by 
\begin{equation}
\begin{split}
\lambda_1&=\frac{a^2+\chi\left(b^2+2\eta ab\cos\theta \right)}{1+\chi \left( 1+4\eta a b \cos \theta \right)},\\
\lambda_2&=1-\lambda_1,\\
\lambda_3&=\lambda_4=0.
\end{split}
\end{equation}
In Fig. \ref{fig:1}, $E^{LT}(\ket{\Psi})$ is plotted  as function of $a^2$ for $\theta=0$ and $\chi=0.3$ for indistinguishable (blue dotted-dotted-dashed line for bosons and red dotted-dashed-dashed line for fermions) and $\chi=0$ for distinguishable particles (black continuous line).
To understand the role played by the localized partial trace and the particle indistinguishability on the entanglement of indistinguishable particles, we compare the results obtained above with the approaches developed in refs. \cite{plastino2009,iemini2013}. In the later case, named second quantization approach, the entanglement measure also employs the von Neumann entropy, but with different features: i) the reduced density matrix is obtained through the trace over the entire degrees of freedom of one particle, which in the present example is $\{ a_{L\uparrow}\ket{0}\!,\!a_{L\downarrow}\ket{0}\!,\! a_{R\uparrow}\ket{0}\!,\!a_{R,\downarrow}\ket{0}\}$; ii) the quantity $\ln N$ is discounted from the entanglement of $N$ particles, so that in this case it becomes $-\ln 2$. Then, the normalized reduced density matrix
\begin{equation}
\label{eq:RDM}
\rho^{SQ}_1=\frac{1}{2\left(\mathcal{N}_1-c_1\right)}\left(
\begin{array}{cccc}
c_1+c_3 & 0 & c_4 & 0 \\
0 & c_1+c_2 & 0 & c_5 \\
c_4^{*} & 0 & c_2 & 0  \\
0 & c_5^{*} & 0 & c_3  
\end{array}
\right).
\end{equation} 
furnishes the following eigenvalues
\begin{equation}
\resizebox{0.43 \textwidth}{!}{$
	\omega_{1}^{\eta}=\omega_{2}^{\eta}= \frac{1}{4} \left\{1+\sqrt{1-\frac{4 a^{2} (1-a^{2}) (1-\chi)^2}{\left[2 \sqrt{a^{2} (1-a^{2})} \chi \cos \theta + \eta \right]^2}}\right\}$}
\end{equation}
and
\begin{equation}
\resizebox{0.43 \textwidth}{!}{$
	\omega_{3}^{\eta}=\omega_{4}^{\eta}=\frac{1}{4} \left\{1-\sqrt{1-\frac{4 a^{2} (1-a^{2}) (1-\chi)^2}{\left[2 \sqrt{a^{2} (1-a^{2})} \chi \cos \theta + \eta \right]^2}}\right\}$},
\end{equation}
which, by its turn, provides the final expression for the entanglement measure 
\begin{equation}
\label{eq:entangSQ}
E^{SQ}(\ket{\Psi})=-\Tr \left( \rho^{SQ}_1 \ln\rho^{SQ}_1 \right) -\ln2=-\sum_{i=1}^{4} \omega^{\eta}_i \ln \omega^{\eta}_i -\ln2.
\end{equation} 
In Fig. \ref{fig:1} $E^{SQ}(\ket{\Psi})$ is plotted as function of $a^2$ for $\theta=0$ and $\chi=0.3$ for indistinguishable particles (blue dotted line for bosons and red dashed line for fermions) and $\chi=0$ for distinguishable particles (black continuous line).

\section{Discussion and final remarks}
\label{sec:Dis}

We have shown that through the second quantization formalism, a very well known tool in many-body physics, together with the localized partial trace over one particle, it is possible to restore the results of the new approach developed by LFC. From the conceptual point view, a state of two particles described according to the second quantization formalism $\ket{\Pi}=a_{j}^{\dagger}a_{k}^{\dagger}\ket{0}$ does not invoke the necessity of symmetrization postulate, since it means the creation of two particles, one in state $j$ and another in state $k$. The symmetrization of the inner product in LFC approach meets its counterpart in the commutation relations between creation and annihilation operators, which define the inner product of states written in second quantization. In other words, once $a_i\ket{0}=0$, it is necessary to perform the normal ordering of creation and annihilation operators to evaluate the expected value of quantum operators composed by creation and annihilation operators in arbitrary order \cite{ballentine2014quantum}.

The impossibility of distinguishing spatially the particles is determined by the parameter $\chi > 0$. In the case in which the particles become distinguishable $\chi=0$, see the black continuous line in Fig. \ref{fig:1}, both measures of entanglement agree, as waited. This property emphasizes the relevance of the term $-\ln 2$ in the measure $E^{SQ}\left(\ket{\Psi}\right)$, which points the necessity of ruling out the contributions coming from the exchange correlations between the particles. This becomes clear for single Slater determinant/permanent states like $\ket{\Pi}=a_{j}^{\dagger}a_{k}^{\dagger}\ket{0}$ in which the von Neumann entropy of the reduced state of one particle is equal to $\ln 2$. 

Another important aspect of the LFC approach is the role played by the localized partial trace. In order to observe its effect we compare the von Neumann entropies of $\rho_1$ and $\rho^{SQ}_1$, or equivalently, $E^{LT}(\ket{\Psi})$ and $E^{SQ}(\ket{\Psi})+\ln 2$, respectively. We notice that the partial trace limits the access to the amount of information of the reduced density matrix of one particle $\rho_1$, and consequently, the amount of entanglement. Additionally, when $E^{LT}(\ket{\Psi})$ \cite{franco2016} is compared to $E^{SQ}(\ket{\Psi})$ \cite{plastino2009,iemini2013}, we observe that the former measure takes in to account contributions coming from exchange correlations between the particles. Therefore, relatively to the second quantization measure of entanglement (\ref{eq:entangSQ}), the LFC approach has a hybrid aspect between entanglement of particles and of modes.

\section*{Acknowledgements} 

A. C. L. and E. I. D. thank to Gabriel T. Landi for valuable discussions. The authors acknowledge the Brazilian National Institute for Science and Technology of Quantum Information and the Brazilian funding agencies CNPq and CAPES for the financial support.

\bibliography{bibliography}

\end{document}